\documentclass[final,3p,times,twocolumn]{elsarticle}
\usepackage{bm}
\usepackage{placeins}
\usepackage{amsfonts}
\usepackage{multirow}
\usepackage{graphicx,color,amsmath,amssymb}
\usepackage[utf8]{inputenc}
\usepackage{verbatim}

\newcommand{\ie}{\textit{i.e.}}

\newcommand{\lsim}{\lesssim}
\newcommand{\gsim}{\gtrsim}

\biboptions{sort&compress}

\begin{document}
\title{Non-equilibrium charmonium regeneration in strongly coupled quark-gluon plasma}

\author{Xiaojian Du$^{1,2}$, Ralf Rapp$^{2}$}
\address{$^1$Fakult\"at fur Physik, Universit\"at Bielefeld, D-33615 Bielefeld, Germany
\\
	$^2$Cyclotron Institute and Department of Physics \& Astronomy, 
	Texas A$\&$M University, College Station, TX 77843-3366, USA
}

\date{\today}

\begin{abstract}
The evaluation of quarkonium regeneration in ultrarelativistic heavy-ion collisions (URHICs) requires the knowledge of the heavy-quark phase space distributions in the expanding quark-gluon plasma (QGP) fireball. We employ a semi-classical charmonium transport approach where regeneration processes explicitly account for the time-dependent spectra of charm quarks via Langevin simulations of their diffusion. 
The inelastic charmonium rates and charm-quark transport coefficients are computed from the same charm-medium interaction. The latter is modeled by perturbative rates, augmented with a $K$-factor to represent nonperturbative interaction strength and interference effect. Using central 5.02~TeV Pb-Pb collisions as a test case we find that a good description of the measured $J/\psi$ yield and its transverse-momentum dependence can be achieved if a large $K\gsim5$ is employed while smaller values lead to marked discrepancies. This is in line with open-charm phenomenology in URHICs, where nonperturbative interactions of similar strength are required. Our approach establishes a common transport framework for a microscopic description of open and hidden heavy-flavor (HF) observables that incorporates both nonperturbative and non-equilibrium effects, and thus enhances the mutual constraints from experiment on the extraction of transport properties of the QGP.
\end{abstract}


\maketitle

\section{Introduction} 
The theoretical description of charmonium production in ultra-relativistic heavy-ion collisions (URHICs) has been investigated for over three decades~\cite{Kluberg:2009wc,BraunMunzinger:2009ih,Rapp:2008tf,Mocsy:2013syh}. 
Early ideas of using the suppression of charmonia as a signature of a deconfined quark-gluon plasma (QGP)~\cite{Matsui:1986dk} have been augmented by the insight that they can also be (re-) generated in the later stages of the fireball evolution, when the temperature has sufficiently dropped for the bound states to re-emerge~\cite{Gazdzicki:1999rk,BraunMunzinger:2000px,Thews:2000rj,Grandchamp:2001pf}. 
The regeneration contribution is then sensitive to the concentration of open-charm states in the system, \ie, $c\bar{c}$ pairs in a deconfined medium or open-charm hadrons in confined matter. 
Initial implementations utilized the statistical hadronization model for the production of all charmonia at the phase boundary~\cite{BraunMunzinger:2000px}, assuming hadronizing charm ($c$) quarks in thermal equilibrium with their number adjusted to hard production in primordial nucleon-nucleon ($NN$) collisions. On the other hand, a gradual screening of the in-medium heavy-quark potential suggests that different charmonium states have different melting temperatures, giving rise to a ``sequential melting". 
This, in turn, implies that also the temperatures at which regeneration occurs are different, \ie, a ``sequential regeneration"~\cite{Du:2015wha}. 

In a dynamical picture, a key role is taken on by the inelastic charmonium reaction rates, which drive the interplay of regeneration and dissociation toward the pertinent equilibrium abundances. This led to the development of transport models including both suppression and regeneration components, applied to URHICs over a large range of collision energies, from the Super-Proton-Synchrotron (SPS) via the Relativistic Heavy-Ion Collider (RHIC) to the Large Hadron Collider (LHC)~\cite{Grandchamp:2003uw,Yan:2006ve,Capella:2007jv,Linnyk:2008hp,Song:2011xi,Chen:2015ona}. 
It was further realized that not only the concentration of open-charm states but also their momentum distributions significantly affect the amount of charmonia regeneration and their transverse-momentum ($p_T$) spectra~\cite{Grandchamp:2002wp,Song:2012at,Zhou:2014kka,He:2021zej}. 
This is so since recombination is most effective from low-momentum $c$ quarks in the thermal rest frame, becoming maximal once they are kinetically equilibrated, while harder spectra, such as those produced in primordial $NN$ collisions, suppress the regeneration rate. 
Similar theoretical efforts have also been undertaken for bottomonium transport in URHICs~\cite{Grandchamp:2005yw,Emerick:2011xu,Strickland:2011aa,Song:2011nu,Krouppa:2015yoa,Du:2017qkv,Chen:2017duy,Brambilla:2017zei,Yao:2020xzw}; some approaches have already included non-equilibrium bottom-quark spectra in the regeneration component, which, however, is quantitatively much smaller than for charmonia and therefore less significant in experimental data. 

In the present paper, we couple a stochastic Langevin simulation of charm-quark spectra to a Boltzmann evolution of charmonia. 
The intimate connection between charm and charmonium transport not only figures through the kinetic off-equilibrium effect in the time-dependent charm spectra, but also via the nonperturbative effects in the basic charm-medium interaction. 
The inelastic reaction rates for charmonia and the transport coefficients for charm quarks share the same collision amplitude~\cite{Du:2019tjf,He:2022ywp}.
Based on our previous charmonium transport calculations~\cite{Du:2015wha}, we adopt perturbative amplitudes for charm-quark scattering off thermal-medium partons, but allow for a phenomenological $K$-factor. 
Recent analyses of open HF transport approaches to experimental observables suggest a $K$-factor of at least 5~\cite{Rapp:2018qla}, highlighting the need for nonperturbative interactions strength commonly believed to be at the core of the strongly-coupled QGP (sQGP).

Our paper is organized as follows: In Sec.~\ref{sec_trans_charmonium}, we briefly introduce the charmonium transport equation and recall its main features as employed previously. 
In Sec.~{\ref{sec_trans_charm}}, we describe our Langevin simulations for $c$ quarks. 
In Sec.~{\ref{sec_coeffs}} we calculate the transport coefficients for elastic $c$-quark medium scattering (Sec.~\ref{ssec_charm-coeff}) and inelastic charmonium reactions (Sec.~\ref{ssec_charmonium-coeff}), 
and highlight their connection. Numerical results for $J/\psi$ $p_T$-spectra are presented in Sec.~\ref{sec_numeric}, and compared to experimental data in Pb-Pb 5\,TeV 
collisions. We conclude in Sec.~\ref{sec_conclusion}.

\section{Charmonium transport approach}
\label{sec_trans_charmonium} 
Our starting point is the Boltzmann equation,
\begin{equation}
\frac{\partial f_\Psi}{\partial t}+\vec{v}\cdot\nabla f_\Psi=-\alpha f_\Psi +\beta \ ,
\label{eq-boltzmann}
\end{equation} 
to describe the evolution of charmonium ($\Psi$=$J/\psi$, $\chi_c$, $\psi'$) phase space distribution functions, $f_\Psi\equiv f_\Psi(\vec{x},\vec{p},t)$ in a fireball, with $\vec{v}$=$\frac{\vec{p}}{E_\Psi(p)}$ and $E_\Psi$=$\sqrt{\vec{p}^2+m_\Psi^2}$; 
$\alpha\equiv\alpha(\vec{p},T(t))$ and $\beta\equiv\beta(\vec{p},T(t))$ denote the dissociation and regeneration rate, respectively. The latter two will be detailed in Sec.~\ref{ssec_charmonium-coeff} below.

The solution to Eq.~(\ref{eq-boltzmann}) can be written as~\cite{Yan:2006ve}
\begin{eqnarray}
f(\vec{x},\vec{p},t)&=&f(\vec{x}-\vec{v}(t-t_0),\vec{p},t_0)
{\rm e}^{-\int_{t_0}^t {\mathrm d}t'\alpha(\vec{p},T(t'))}
\nonumber\\
&+&\int_{t_0}^t {\mathrm d}t'\beta(\vec{p},T(t'))
{\rm e}^{-\int_{t'}^t {\mathrm d}t''\alpha(\vec{p},T(t''))} \  .
\label{eq-bsol}
\end{eqnarray} 

The first term on the right-hand-side ($rhs$) describes the dissociation of primordially produced charmonia. 
For their initial distribution, we adopt a factorized ansatz $f(\vec{x},\vec{p},t_0,b)=f(\vec{x},b)f(\vec{p})S(b)$, with a spatial distribution, $f(\vec{x},b)$, evaluated from a collision profile from a Glauber model~\cite{Miller:2007ri} ($b$: impact parameter), and a momentum spectrum, $f(\vec{p})$, based on experimental data in $pp$ collisions~\cite{Bossu:2011qe,Book:2015}. 
The latter is modified by a factor, $S(b)$, associated with cold-nuclear-matter effects in the incoming nuclei, \ie,  
shadowing of the nuclear parton distribution functions, an effective nuclear absorption cross section (mostly relevant at lower collisions energies), and a Cronin effect~\cite{Gavin:1988tw,Hufner:1988wz} through a Gaussian smearing of the
$pp$ spectra. 
We have constrained these effects in our previous study of
charmonium production in d-Au collisions at RHIC and $p$-Pb collisions at the LHC~\cite{Du:2018wsj}. The suppression effect due to the hot QCD medium is encoded in the exponential factors on the $rhs$ of Eq.~(\ref{eq-bsol}) through the dissociation rate  $\alpha(\vec{p},T)$.

The second term on the $rhs$ of Eq.~(\ref{eq-bsol}) accounts for charmonium regeneration, which, in turn, is also subject to suppression from time $t'$ on. 
These processes, which are controlled by the regeneration rate, $\beta(\vec{p},T)$, are due to recombination of $c$ and $\bar c$ quarks and thus depend on their individual momentum spectra. 
In particular, the softening of the initial power-law $c$-quark spectra due to their diffusion through the medium enhances the phase-space overlap for charmonium formation. 
In our previous work within the kinetic-rate equation~\cite{Grandchamp:2002wp}, we have approximated this effect by a thermal relaxation factor, while the $p_T$ spectra
were approximated by a thermal blastwave model~\cite{Zhao:2007hh,Du:2015wha}. 
In the present work we lift these approximations by simulating the time dependence of the $c$-quark spectra via their transport through the QGP and injecting them into the charmonium Boltzmann equation.
 
\section{Charm-quark transport}
\label{sec_trans_charm} 
The $c$-quark diffusion is described by a stochastic Langevin equation~\cite{He:2013zua},
\begin{align}
{\mathrm d}\vec{x}=\frac{\vec{p}}{E_c}{\mathrm d}t \ , \ 
{\mathrm d}\vec{p}=-A\vec{p}{\mathrm d}t+\sqrt{2{\mathrm d}t D}\vec{\rho} \ 
\label{eq-langevin}
\end{align}
with $E_c$=$\sqrt{\vec{p}^2+m_c^2}$. The momentum change is due to a friction term causing energy loss at a rate $A$$\equiv$$A(\vec{p},T(t))$, and a diffusion term with coefficient $D$$\equiv$$D(\vec{p},T(t))$, randomized by a Gaussian noise $\vec{\rho}$. 
The corresponding Kolmogorov forward-equation is the Fokker-Planck equation, 
\begin{equation}
\frac{\partial}{\partial t}f_c(\vec{p},t)
=\nabla_{p}\left(A\vec{p}f_c(\vec{p},t)\right)
+\nabla_{p}^2\left(Df_c(\vec{p},t)\right) \ ,
\end{equation}
for the $c$-quark distribution, $f_c$.
Its stationary solution is the Boltzmann distribution, $f^{\rm eq}(\vec{p},T(t))\propto {\rm exp}\left(-\frac{E_c}{T}\right)$, implying an Einstein relation
\begin{equation}
A(\vec{p},T(t))=\frac{1}{E_c}\left(\frac{D(\vec{p},T(t))}{T}
-\frac{\partial D(\vec{p},T(t))}{\partial E_c}\right) \ .
\end{equation}
The relaxation rate, $A(\vec{p},T)$, is calculated from the $c$-quark amplitude for elastic scattering off medium partons, cf.~Sec.~\ref{ssec_charm-coeff} below.
With $m_c\gg T$, the derivative term is subleading so that one can approximate the diffusion coefficient as~\cite{He:2013zua}
\begin{equation}
D(\vec{p},T(t))\approx A E_c T(t) \ . 
\end{equation}
The stochastic simulation evolves the $c$-quark spectra toward their local equilibrium distribution. 
The initial momenta are sampled according to primordial production in $pp$ collisions, fitted to pQCD results in the Fixed-Order Next-to-Leading-Logarithm (FONLL) scheme~\cite{Cacciari:2012ny}. 
The initial spatial distribution is uniformly sampled.

\section{Transport coefficients}
\label{sec_coeffs}
In this section we elaborate on the transport coefficients for $c$-quark
diffusion (Sec.~\ref{ssec_charm-coeff}) and charmonium kinetics (Sec.~\ref{ssec_charmonium-coeff}). 
For simplicity, we adopt pQCD Born amplitudes including a $K$-factor to mimic nonperturbative effects, augmented with an additional interference factor for charmonium states~\cite{Laine:2006ns,Liu:2017qah}.

\subsection{Relaxation rate for charm quarks}
\label{ssec_charm-coeff}
\begin{figure}[!t]
	\begin{tabular}{c}
		\begin{minipage}[b]{0.95\linewidth}
			\includegraphics[width=1.0\textwidth]{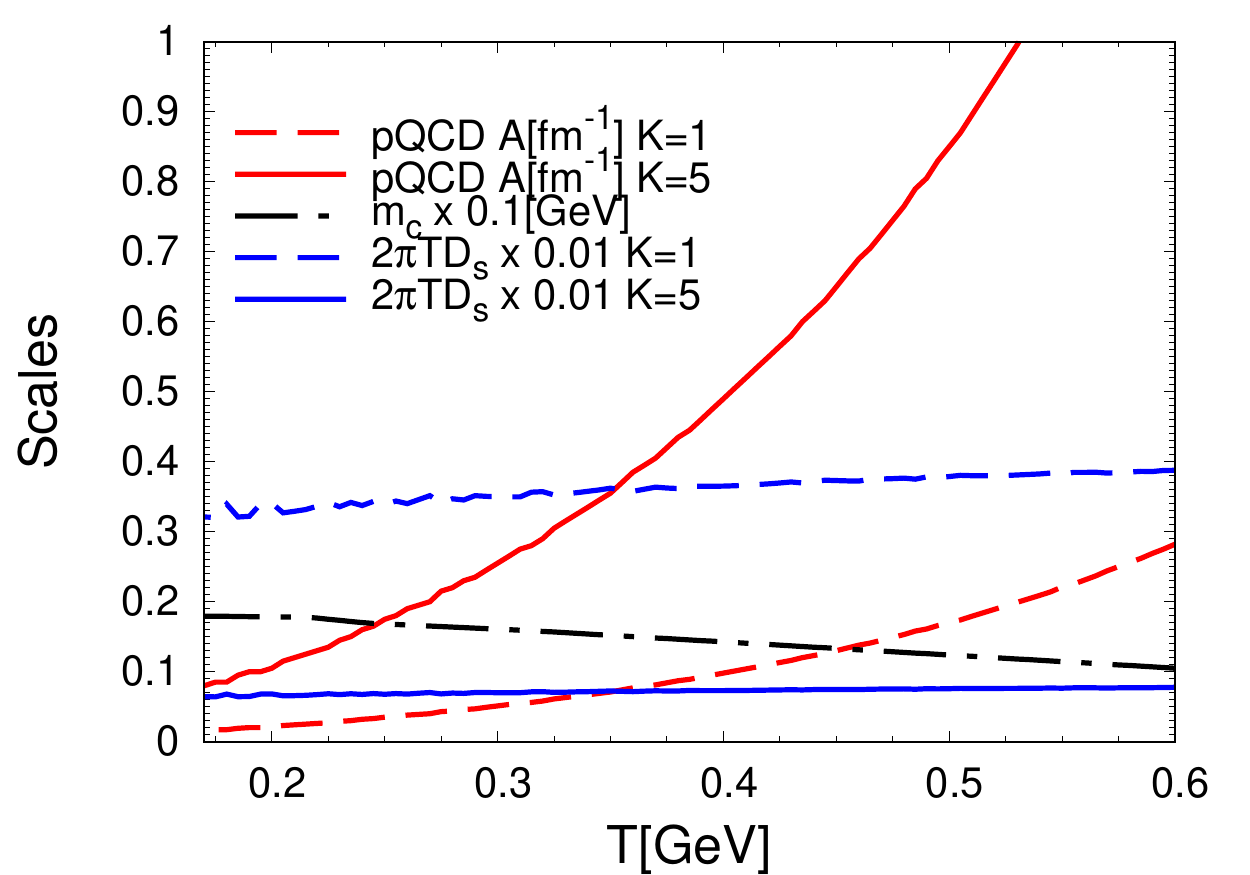}
			\caption{Relaxation rates for $c$ quarks using pQCD Born amplitudes with coupling constant $g$=2 for $K$=1 (dashed red) and $K$=5 (solid red) with $c$-quark mass $m_c(T)$ (dotted-dashed black), and the corresponding spatial diffusion coefficient ${\cal D}_s(2\pi T)$=$\frac{2\pi T^2}{m_c A(p=0)}$ for $K$=1 (dashed blue) and $K$=5 (solid blue)}
			\label{fig_coeff}
		\end{minipage}
	\end{tabular}
\end{figure}

The relaxation rate $A$ in Eq.~(\ref{eq-langevin}) can be written in terms of the 2$\rightarrow$2 scattering amplitude, $M_{ic\rightarrow{i}c}$, of $c$ quarks on thermal partons ($i=q, \bar{q}, g$) as~\cite{Svetitsky:1987gq}
\begin{align}
&A(\vec{p}_c,T)=\sum_i\int {\mathrm d}\Pi_{\rm c} \ 
d_{i} \ \overline{|M_{ic\rightarrow ic}|^2}
\nonumber\\
&\quad \times
[1\pm f_{i}(\tilde{p}_{i})] \  [1-f_{c}(p_c')]
f_{i}(p_i)\left(1-\frac{\vec{p}_c\cdot\vec{p}_c'}{\vec{p}_c^2}\right)
\label{eq-relaxation}
\end{align}
with the measure
\begin{align}
\nonumber
{\mathrm d}\Pi_{\rm c}&=\frac{1}{2E_{\Psi}}
\frac{{\mathrm d}^3p_i}{(2\pi)^{3}2E_i}
\frac{{\mathrm d}^3\tilde{p}_i}{(2\pi)^{3}2\tilde{E}_i}
\frac{{\mathrm d}^3p_c'}{(2\pi)^{3}2E_c'}\\
&\quad \times(2\pi)^{4}\delta^{(4)}\left(p_c+p_i-\tilde{p}_i-p_c'\right) \ , 
\label{eq-measure2}
\end{align}
where $\vec{p}_c(\vec{p}_c')$ and $\vec{p}_i(\vec{\tilde{p}}_i)$ denote the incoming (outgoing) momenta of $c$ quarks and light partons, respectively ($f_i$: thermal-parton distributions). 
For the matrix element we use pQCD Born diagrams with a Debye mass, $m_D=gT$, to regulate the $t$-channel gluon exchange, and a coupling constant $g$=2. 
As is well known, the pertinent diffusion coefficient of ${\cal D}_s(2\pi T)$$\sim$30 is much too large to account for low-$p_T$ HF phenomenology in URHICs~\cite{Rapp:2018qla}; we therefore allow for an overall $K$ factor multiplying $A(p)$, see Fig.~\ref{fig_coeff}. HF phenomenology suggests $K\gtrsim 5$ which results in a spatial diffusion coefficient ${\cal D}_s(2\pi T)\lesssim 6$.

\subsection{Quasi-free reaction rates for charmonia}
\label{ssec_charmonium-coeff}
For moderate binding energies, $E_B\lsim T$, inelastic scattering, $i+\Psi\to i+c+{\bar c}$ ($i=g,q,\bar{q}$), is dominant over gluo-dissociation processes, $g+\Psi \to c +{\bar c}$~\cite{Grandchamp:2001pf,Zhao:2010nk}. 
Thus, the charmonium reaction rates in Eq.~(\ref{eq-bsol}) take the form
\begin{align}
&\alpha(\vec{p},T)=\sum_i\int {\mathrm d}\Pi_{\rm \Psi} \ 
d_{i} \ \overline{|M_{i\Psi\rightarrow \tilde{i}c\bar{c}}|^2}
\nonumber\\
& \qquad \times [1\pm f_{i}(\tilde{p}_{i})] [1-f_{c}(p_{c})] [1-f_{\bar{c}}(p_{\bar{c}})] f_{i}(p_i) \ ,
\label{eq-alpha}
\\
&\beta(\vec{p},T)=\sum_i\int {\mathrm d}\Pi_{\rm \Psi} \ 
d_{\tilde{i}} \ d_{c} \ d_{\bar{c}} \ \overline{|M_{\tilde{i}c\bar{c}\rightarrow i\Psi}|^2}
\nonumber \\
& \qquad \times [1\pm f_{i}(p_i)] f_{i}(\tilde{p}_{i}) \gamma_c f_{c}(p_{c}) \gamma_c f_{\bar{c}}(p_{\bar{c}}) \ ,
\label{eq-beta}
\end{align}
with parton degeneracies $d_{c,i}$ and the measure
\begin{align}
\nonumber
{\mathrm d}\Pi_{\rm \Psi}=&\frac{1}{2E_{\Psi}}
\frac{{\mathrm d}^3p_i}{(2\pi)^{3}2E_i}
\frac{{\mathrm d}^3\tilde{p}_i}{(2\pi)^{3}2\tilde{E}_i}
\frac{{\mathrm d}^3p_c}{(2\pi)^{3}2E_c}
\frac{{\mathrm d}^3p_{\bar{c}}}{(2\pi)^{3}2E_{\bar{c}}}
\nonumber\\ 
&  \times(2\pi)^{4}\delta^{(4)}\left(p+p_i-\tilde{p}_i-p_c-p_{\bar{c}}\right).
\label{eq-measure}
\end{align}
Detailed balance requires the relation
\begin{eqnarray}
d_{i}d_{\Psi}\overline{|M_{i\Psi\rightarrow \tilde{i}c\bar{c}}|^2} =
d_{\tilde{i}}d_{c}d_{\bar{c}}\overline{|M_{\tilde{i}c\bar{c}\rightarrow i\Psi}|^2} \ .
\end{eqnarray}
In equilibrium, the reaction rates are related as
\begin{eqnarray}
\beta(\vec{p},T)=\gamma_c^2d_{\Psi}{\rm exp}\left(-\frac{E_{\Psi}}{T}\right)\alpha(\vec{p},T) \ , 
\end{eqnarray}
where the fugacity factor, $\gamma_c$=$\gamma_{\bar c}$, of anti-/charm quarks accounts for their number conservation.

As in our previous works, based on the relatively small charmonium binding energies and the notion that $c\bar c$ final-state interactions in the color-octet channel are 1/$N_c^2$ suppressed ($N_c$=3), we employ a ``quasifree" approximation to compute the inelastic 3$\leftrightarrow$2 amplitudes. 
This amounts to factorizing the latter into elastic 2$\leftrightarrow$2 ones for scattering thermal partons off individual $c$ and $\bar c$ quarks in the bound state while conserving 4-momentum, 
\begin{align}
&\overline{|M_{i\Psi\rightarrow \tilde{i}c\bar{c}}(p_i,p_{\Psi},{\tilde p}_i,p_c,p_{\bar{c}})|^2} \qquad
\nonumber \\
&\qquad =2\overline{\left|M_{ic\rightarrow \tilde{i}c}\left(p_i,\frac{m_{\Psi}-m_c}{m_{\Psi}}p_{\Psi},{\tilde p}_i,p_c\right)\right|^2}
\nonumber\\
& \quad \qquad \times(2\pi)^3 2E_{\bar{c}}\delta^{(3)}\left(\vec{p}_{\bar{c}}-\frac{m_c}{m_{\Psi}}\vec{p}_{\Psi}\right) \ .
\label{eq-reduction}
\end{align}
In doing so, the binding energy of the $\Psi$ is included in the mass of the incoming $c$ quark participating in the scattering, $m_c \to m_c-E_B$, while the $\bar c$ quark acts as a spectator, which implies incoming 3-momenta as 
\begin{align}
p^{in}_{c}=\frac{m_{\Psi}-m_c}{m_{\Psi}}p_{\Psi} \ , \quad p_{\bar{c}}^{in}=\frac{m_c}{m_{\Psi}}p_{\Psi} \ .
\end{align}
Since the spectator's momentum does not change, the momentum conservation in Eq.~(\ref{eq-measure}) becomes
\begin{align}
\nonumber
\delta^{(4)}(p_{\Psi}+p_i-{\tilde p}_i-p_{c}-p_{\bar{c}})
=\delta^{(4)}\left(p_c^{in}+p_i-{\tilde p}_i-p_{c}\right) \  ,
\end{align}
and the final spectator momentum integrates out as 
\begin{equation}
\int \frac{d^3p_{\bar{c}}}{(2\pi)^3 2E_{\bar{c}}}(2\pi)^3 2E_{\bar{c}}\delta^{(3)}\left(\vec{p}_{\bar{c}}-\frac{m_c}{m_{\Psi}}\vec{p}_{\Psi}\right)=1 \ .
\end{equation}
This is the origin of the extra factor in Eq.~(\ref{eq-reduction}), while an additional factor of 2 accounts for inelastic scattering off the $\bar c$ quark.
Inserting this into Eqs.~(\ref{eq-alpha}) and (\ref{eq-beta}) leads to the following reaction rates:
\begin{align}
&\alpha(\vec{p}_{\Psi},T)=2d_{i}\sum_i\int {\mathrm d}\Pi_{\rm QF}
\overline{\left|M_{ic\rightarrow \tilde{i}c}\left(p_i,p_c^{in},{\tilde p}_i,p_c\right)\right|^2}
\nonumber\\
&\times [1\pm f_{i}(\tilde{p}_{i})] \ [1-f_{c}(p_{c})] \ [1-f_{\bar{c}}(\frac{m_c}{m_{\Psi}}p_{\Psi})] 
\ f_{i}(p_i)
\label{eq-alpha-qf}
\\
&\beta(\vec{p}_{\Psi},T)=2\gamma_c^2 d_{\Psi}d_{i}\sum_i\int {\mathrm d}\Pi_{\rm QF}
\overline{\left|M_{ic\rightarrow \tilde{i}c}\left(p_i,p_c^{in},{\tilde p}_i,p_c\right)\right|^2}
\nonumber \\
&\times
[1\pm f_{i}(p_i)] \ 
f_{i}(\tilde{p}_{i}) \ 
f_{c}(p_{c}^{in}) \
f_{\bar{c}}(\frac{m_c}{m_{\Psi}}p_{\Psi})
\label{eq-beta-qf}
\end{align}
with the quasifree measure
\begin{align}
\nonumber
{\mathrm d}\Pi_{\rm QF}&=\frac{1}{2E_{\Psi}}
\frac{{\mathrm d}^3p_i}{(2\pi)^{3}2E_i}
\frac{{\mathrm d}^3\tilde{p}_i}{(2\pi)^{3}2\tilde{E}_i}
\frac{{\mathrm d}^3p_c}{(2\pi)^{3}2E_c}\\
& \ \  \times(2\pi)^{4}\delta^{(4)}\left(\frac{m_{\Psi}-m_c}{m_{\Psi}}p_{c}^{in}+p_i-\tilde{p}_i-p_c\right).
\label{eq-measure-qf}
\end{align}
One can now appreciate the explicit relation between the quasifree charmonium reaction rates, $\alpha$ and $\beta$, and the relaxation rate $A$ for charm quarks, in terms of the same underlying heavy-light scattering amplitude. 
In particular, nonperturbative effects, which we mimic by a $K$-factor, figure on equal footing for phenomenological extractions of heavy-quark and quarkonium transport parameters. 
Finally, we account for a suppression in the charmonium rates due to the interference of the amplitudes for the $c$-quark and antiquark (which, in particular, ensures that for vanishing bound-state size the rate becomes zero). Toward this end we augment the charmonium rates, $\alpha$ and $\beta$, by an interference factor inspired by the pQCD form~\cite{Laine:2006ns}
\begin{eqnarray}
\phi(x)=2\int_0^{\infty}\frac{zdz}{(z^2+1)^2}\left(1-\frac{\sin(zx)}{zx}\right) \ ,
\end{eqnarray}
where $x=m_Dr$ and $r$ is the average radius of the charmonium state. We approximate the radius by an inverse relation to the charmonium binding energy $r\simeq \frac{C\alpha_s}{E_B}$~\cite{Du:2017qkv}, where  $C\simeq 3-4$ is estimated from the vacuum values.

The time-dependent $c$-quark distributions,  $f_{c,\bar c}(\vec{p}_{c,\bar c})$ directly control the charmonium regeneration rate $\beta(\vec{p},T(t))$. 
For quantitative applications one needs to determine their absolute norm, which is set by their initial hard production in binary $NN$ collisions (including shadowing corrections) and assumed to be conserved thereafter. 
This is done in the same way as in the statistical hadronization model, by introducing a fugacity factor 
$\gamma_c$=$\gamma_{\bar{c}}$
according to
\begin{equation}
N_{c\bar{c}}=\frac{1}{2}\gamma_c(t) n_{c}(T(t))V_{\rm FB}(t)\frac{I_{1}(\gamma_c(T) n_{c}(T(t))V_{\rm FB}(t))}{I_{0}(\gamma_c(t) n_{c}(T(t))V_{\rm FB}(t))},
\label{eq_fugacity}
\end{equation}
where the total $c$+$\bar{c}$ density reads,
\begin{equation}
n_{c}(T(t))=2d_c\int\frac{\mathrm{d}^3p}{(2\pi)^3}f_{c}(\vec{p},T(t)) \ , 
\end{equation}
are computed with their thermal off-equilibrium distributions at each time and temperature from the Langevin simulation discussed in~Sec.~\ref{sec_trans_charm}.

When the fireball approaches and reaches the critical temperature, charmonium regeneration from charm hadrons comes into play. The explicit treatment of these processes is beyond the scope of the present study, and therefore we implement them in a simplified form through the mixed phase at $T_c=180$\,MeV in our fireball (corresponding to the time window from $\sim$5.5-9\,fm/c in Fig.~\ref{fig_trial-A}).
Noting that the fugacity factor is inversely proportional to the open-charm density at fixed temperature, $\gamma_c\sim \frac{1}{n_c}$, and that the regeneration rate quadratically depends on fugacity factor, $\beta\sim\gamma_c^2 \alpha$ (in relaxation time approximation, $\beta = \alpha N_\Psi^{\rm eq}$), we can approximate the hadronic regeneration rate at $T_c$ as
\begin{equation}
\beta^{\rm had}(\vec{p},T)\simeq \beta^{\rm qgp}(\vec{p},T) \frac{\alpha^{\rm had}(\vec{p},T)}{\alpha^{\rm qgp}(\vec{p},T)}  \left(\frac{n_c^{\rm qgp}}{n_c^{\rm had}}\right)^2 \ ,
\end{equation}
where the superscript ``had" refers to the quantities in hadronic matter taken from Ref.~\cite{Du:2015wha}. We then employ the standard mixed-phase construction to partition the hadronic and QGP quantities according to
\begin{eqnarray}
X^{\rm mix}(t) =f(t)X^{\rm qgp}(T_c)+(1-f(t))X^{\rm had}(T_c) \ 
\end{eqnarray}
($X$=$\alpha$, $\beta$), where $f(t)$ denotes the volume fraction of the mixed phase occupied by QGP (as inferred from entropy conservation).

\section{$J/\psi$ yields and spectra in URHICs}
\label{sec_numeric}
We now turn to the numerical results of the coupled Langevin-Boltzmann framework for $c$ and $\Psi$ transport within the expanding fireball framework for URHICs as developed and deployed in our previous works.

\begin{figure}
	\begin{tabular}{c}
		\begin{minipage}[b]{0.95\linewidth}
			\includegraphics[width=1.0\textwidth]{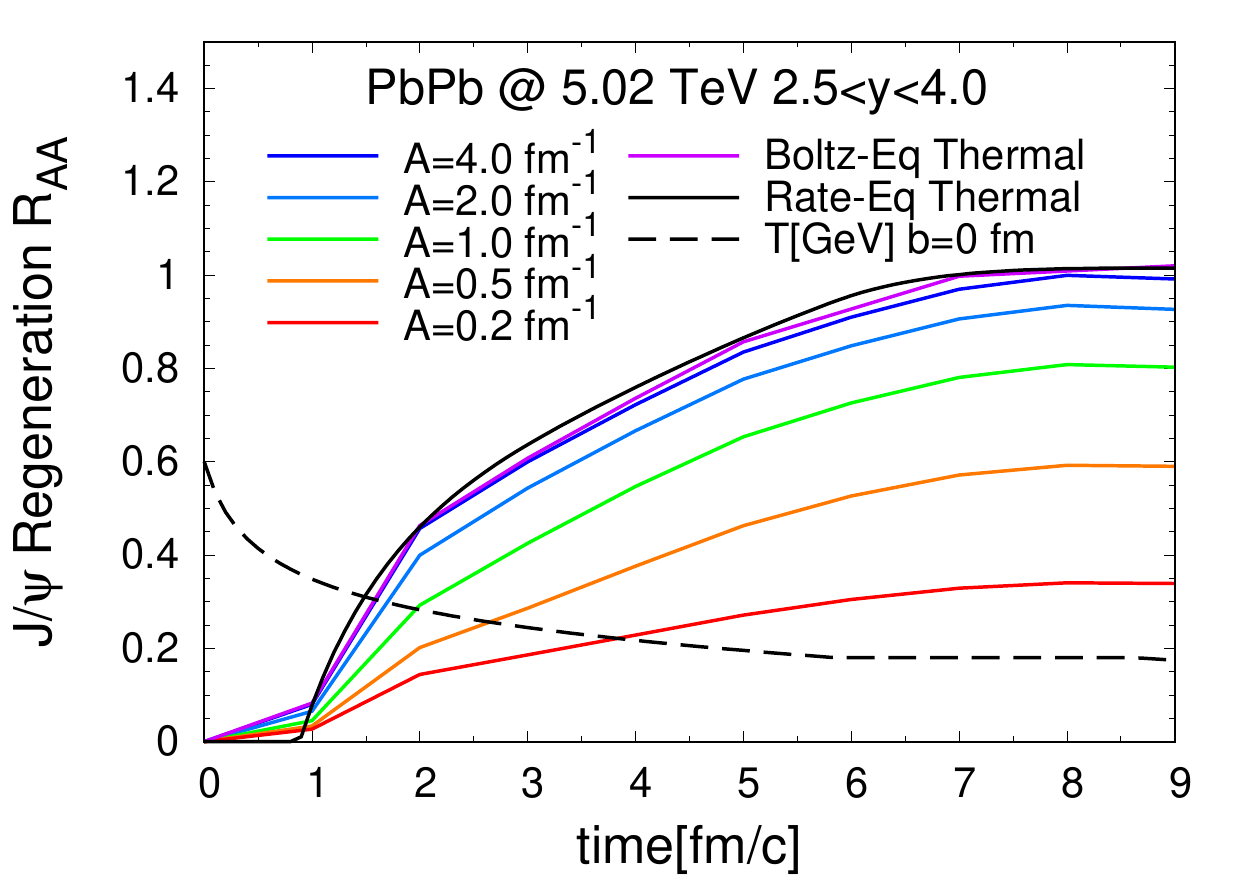}
\caption{Results from our coupled Langevin-Boltzmann approach for the regeneration component of the 
direct-$J/\psi$ $R_{\rm AA}(t)$ in central Pb-Pb collisions for constant c-quark relaxation rates, $A$=0.2,0.5,1,2,4\,fm$^{-1}$  
(red, orange, green, light-blue, and blue lines, respectively), and for (instantly) equilibrated c-quark spectra (purple line). Also shown are the equilibrium limit from our previous rate equation approach (black line), and the temperature evolution until the end of the QGP phase (dashed black line). 
}
\label{fig_trial-A}
		\end{minipage}
	\end{tabular}
\end{figure}
First we conduct a check of the thermal equilibrium limit, by carrying out Langevin simulations for $c$-quarks with fixed thermalization rates between $A$=0.2-4\,fm$^{-1}$ and coupling these into the $J/\psi$ Boltzmann equation using the basic pQCD matrix element (no $K$ factor) for the reaction rates in central 5\,TeV Pb-Pb collisions. 
The resulting time dependence of the regeneration component of the direct-$J/\psi$ $R_{\rm AA}$ (without feeddown) shows the expected increase with increasing $A$ and converges to the equilibrium limit obtained from thermal $c$-quark distributions (and as previously used in our rate-equation approach), cf.~Fig.~\ref{fig_trial-A}. 
We also confirm that the regeneration yield in off-equilibrium scenarios, say, for $A$=0.5\,fm$^{-1}$, is quite consistent with a schematic relaxation time correction factor employed before~\cite{Grandchamp:2002wp,Zhao:2010nk,Song:2012at}.

\begin{figure}[!t]
	\begin{tabular}{c}
		\begin{minipage}[b]{0.95\linewidth}
			\includegraphics[width=1.0\textwidth]{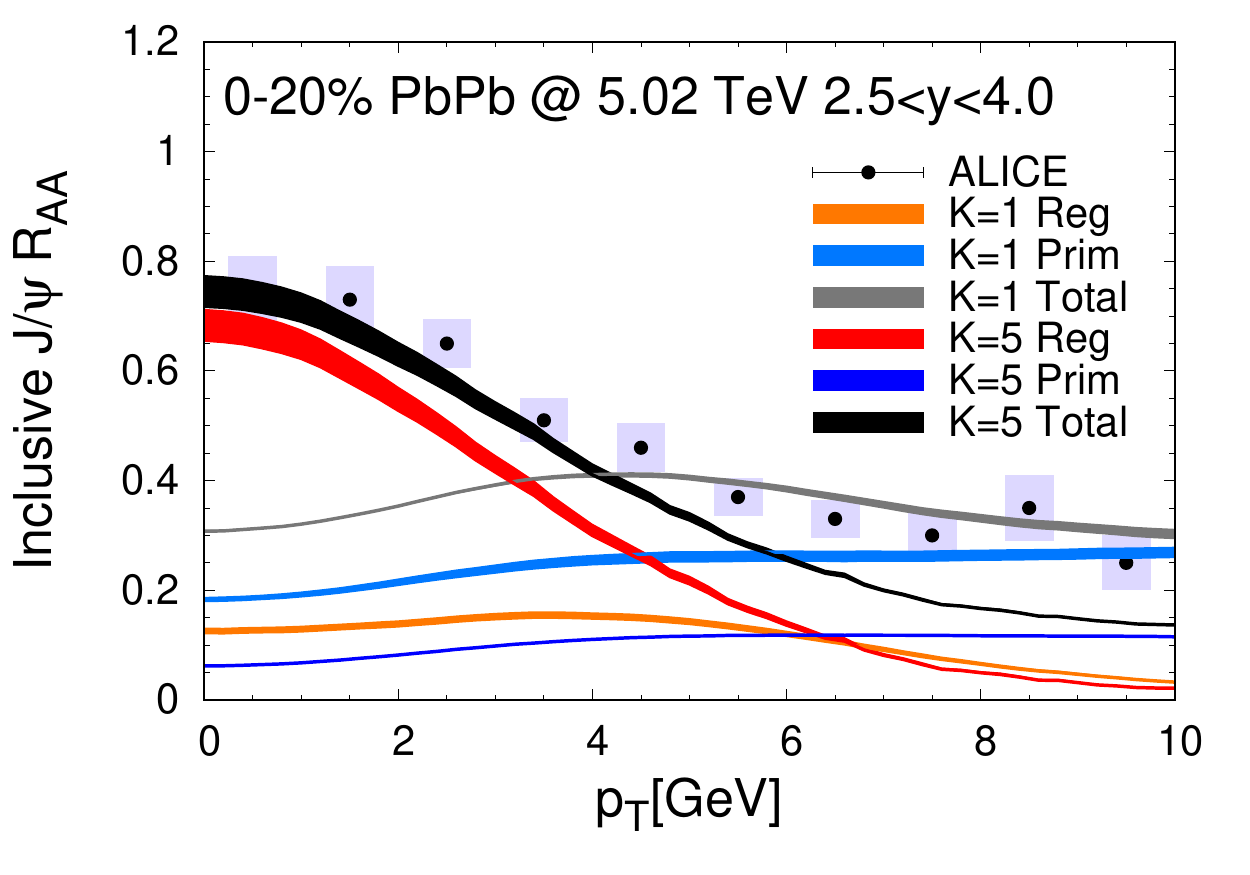}
\caption{Inclusive $J/\psi$ $R_{\rm AA}(p_T)$ calculated from Langevin-Boltzmann transport for charm/onia
using the same pQCD matrix elements with different $K$ factors of 1 (orange, light blue and grey bands for regeneration, primoridial and total yields, respectively) and 5 (red, blue, and black bands), compared to ALICE data~\cite{Adam:2016rdg}.}
			\label{bolt-exp}
		\end{minipage}
	\end{tabular}
\end{figure}

Next, we perform our microscopic simulations, focusing on the resulting $p_T$ spectra plotted in terms of the inclusive-$J/\psi$ $R_{\rm AA}(p_T)$ in central Pb-Pb (5\,TeV) collisions at forward rapidity (including feeddown from $\chi_c$ and 
$\psi'$ states, as well as $b$ decays), see Fig.~\ref{bolt-exp}.  
For $K$=1 the $c$-quark relaxation rates are clearly too small, generating too little yield and too hard spectra in comparison to ALICE data~\cite{Adam:2016rdg}. However, for $K$=5 a fair description of both total yield and spectral shape at low and intermediate $p_T$ is found. This is rather remarkable, since it is consistent with independently inferred $K$ factors from (a) low-$p_T$ open-charm observables~\cite{Rapp:2018qla}, and (b) a recent statistical extraction of bottomonium data at RHIC and the LHC~\cite{Du:2019tjf} (where regeneration plays a subleading role). For $p_T\gsim6$\,GeV the ALICE data are underestimated, but this is not unreasonable as one expects the $K$ factor to decrease with momentum. Non-thermal anisotropic charm spectra have also been found to be critical for the description of $J/\psi$ elliptic flow for $p_T$$\simeq$4-8\,GeV~\cite{He:2021zej,He:2022ywp}.

The most significant difference in the outcome of our coupled transport approach, relative to our previous rate equation results~\cite{Grandchamp:2001pf,Zhao:2010nk}, is the magnitude of the charmonium reaction rates. The $K$ factor renders them significantly larger than before, due to the much larger interaction strength which is now compatible with the notion of a strongly coupled QGP (sQGP).

\section{Conclusions}
\label{sec_conclusion} 
We have constructed a coupled transport approach of open and hidden charm within a Langevin-Boltzmann framework, by implementing explicitly evolving $c$-quark distributions into charmonium kinetics.
Utilizing a quasifree approximation, as appropriate for the relatively weak binding of the $J/\psi$ (and excited states) in the QGP, the same charm-medium scattering amplitude figures in both open and hidden charm transport equations, thus manifesting the connection between the medium properties of the two sectors.

In a first application, and to make contact with our previous rate equation results, we have employed pQCD Born amplitudes for the $c$-quark interaction with thermal-medium partons. 
The simulated $c$-quark spectra remain rather hard throughout the QGP evolution, providing little phase space overlap for charmonium regeneration. 
The resulting $J/\psi$ yields are too small, and their $p_T$ spectra too hard, in comparison to experimental data in Pb-Pb collisions at the LHC, describing the data only toward higher $p_T$. 
However, when augmenting the (squared) amplitude with an overall $K$-factor of 5, emulating nonperturbative interaction strength, a simultaneous description of the measured yields and $p_T$ spectra at low and intermediate $p_T$ for inclusive $J/\psi$ production in central Pb-Pb collisions is achieved. 
We thus realize a nonperturbative description of $J/\psi$ observables characteristic for a sQGP, with a medium coupling comparable to recent extractions from bottomonium 
suppression and open HF diffusion calculations. At higher $p_T$, the data are better described with smaller $K$ factors, which is consistent with the generic momentum dependence of the QCD coupling strength.

Our results warrant a systematic re-investigation of charmonium data at SPS, RHIC and the LHC. In particular, the schematic $K$ factor should be replaced by explicit nonperturbative calculations, such as those from the $T$-matrix approach which have been constrained by a variety of lattice-QCD ``data".  Another improvement concerns the inclusion of quantum effects in the transport equations in a more explicit manner. 
In the quarkonium sector, several efforts in this direction are in progress~\cite{Akamatsu:2011se,Borghini:2011ms,Brambilla:2016wgg,Blaizot:2018oev,Katz:2015qja,Chen:2016vha,Yao:2018nmy,Boyd:2019arx,Brambilla:2020qwo}. 
Together, these developments can be expected to advance the understanding of the sQGP through HF observables in URHICs to the next level.

\section*{Acknowledgments}
We thank M.~He and S.Y.F~Liu for valuable discussion. This work has been supported by the U.S. National Science Foundation under grant no.~PHY-1913286 (RR); XD acknowledges support by the Deutsche
Forschungsgemeinschaft (DFG) through grant CRC-TR 211 
project no.~315477589-TRR 211, and computing time provided by the Paderborn Center for Parallel Computing (PC2).

\bibliographystyle{h-elsevier}

\bibliography{refnoda}

\begin{thebibliography}{10}

\bibitem{Kluberg:2009wc}
L. Kluberg and H. Satz,
\newblock Landolt-Bornstein 23 (2010) 372.

\bibitem{BraunMunzinger:2009ih}
P. Braun-Munzinger and J. Stachel,
\newblock Landolt-Bornstein 23 (2010) 424.

\bibitem{Rapp:2008tf}
R. Rapp, D. Blaschke and P. Crochet,
\newblock Prog. Part. Nucl. Phys. 65 (2010) 209.

\bibitem{Mocsy:2013syh}
A. Mocsy, P. Petreczky and M. Strickland,
\newblock Int. J. Mod. Phys. A 28 (2013) 1340012.

\bibitem{Matsui:1986dk}
T. Matsui and H. Satz,
\newblock Phys. Lett. B178 (1986) 416.

\bibitem{Gazdzicki:1999rk}
M. Gazdzicki and M.I. Gorenstein,
\newblock Phys. Rev. Lett. 83 (1999) 4009.

\bibitem{BraunMunzinger:2000px}
P. Braun-Munzinger and J. Stachel,
\newblock Phys. Lett. B490 (2000) 196.

\bibitem{Thews:2000rj}
R.L. Thews, M. Schroedter and J. Rafelski,
\newblock Phys. Rev. C63 (2001) 054905.

\bibitem{Grandchamp:2001pf}
L. Grandchamp and R. Rapp,
\newblock Phys. Lett. B523 (2001) 60.

\bibitem{Du:2015wha}
X. Du and R. Rapp,
\newblock Nucl. Phys. A943 (2015) 147.

\bibitem{Grandchamp:2003uw}
L. Grandchamp, R. Rapp and G.E. Brown,
\newblock Phys. Rev. Lett. 92 (2004) 212301.

\bibitem{Yan:2006ve}
L. Yan, P. Zhuang and N. Xu,
\newblock Phys. Rev. Lett. 97 (2006) 232301.

\bibitem{Capella:2007jv}
A. Capella et~al.,
\newblock Eur. Phys. J. C58 (2008) 437.

\bibitem{Linnyk:2008hp}
O. Linnyk, E.L. Bratkovskaya and W. Cassing,
\newblock Int. J. Mod. Phys. E17 (2008) 1367.

\bibitem{Song:2011xi}
T. Song, K.C. Han and C.M. Ko,
\newblock Phys. Rev. C 84 (2011) 034907.

\bibitem{Chen:2015ona}
B. Chen, P. Zhuang and Z. Xu,
\newblock Phys. Rev. C93 (2016) 044917.

\bibitem{Grandchamp:2002wp}
L. Grandchamp and R. Rapp,
\newblock Nucl. Phys. A709 (2002) 415.

\bibitem{Song:2012at}
T. Song, K.C. Han and C.M. Ko,
\newblock Phys. Rev. C85 (2012) 054905.

\bibitem{Zhou:2014kka}
K. Zhou et~al.,
\newblock Phys. Rev. C89 (2014) 054911.

\bibitem{He:2021zej}
M. He, B. Wu and R. Rapp,
\newblock Phys. Rev. Lett. 128 (2022) 162301.

\bibitem{Grandchamp:2005yw}
L. Grandchamp et~al.,
\newblock Phys. Rev. C 73 (2006) 064906.

\bibitem{Emerick:2011xu}
A. Emerick, X. Zhao and R. Rapp,
\newblock Eur. Phys. J. A48 (2012) 72.

\bibitem{Strickland:2011aa}
M. Strickland and D. Bazow,
\newblock Nucl. Phys. A 879 (2012) 25.

\bibitem{Song:2011nu}
T. Song, K.C. Han and C.M. Ko,
\newblock Phys. Rev. C85 (2012) 014902.

\bibitem{Krouppa:2015yoa}
B. Krouppa, R. Ryblewski and M. Strickland,
\newblock Phys. Rev. C 92 (2015) 061901.

\bibitem{Du:2017qkv}
X. Du, R. Rapp and M. He,
\newblock Phys. Rev. C 96 (2017) 054901.

\bibitem{Chen:2017duy}
B. Chen and J. Zhao,
\newblock Phys. Lett. B 772 (2017) 819.

\bibitem{Brambilla:2017zei}
N. Brambilla et~al.,
\newblock Phys. Rev. D97 (2018) 074009.

\bibitem{Yao:2020xzw}
X. Yao et~al.,
\newblock JHEP 01 (2021) 046, 2004.06746.

\bibitem{Du:2019tjf}
X. Du, S.Y.F. Liu and R. Rapp,
\newblock Phys. Lett. B796 (2019) 20.

\bibitem{He:2022ywp}
M. He, H. van Hees and R. Rapp,
\newblock (2022), 2204.09299.

\bibitem{Rapp:2018qla}
A. Beraudo et~al.,
\newblock Nucl. Phys. A979 (2018) 21.

\bibitem{Miller:2007ri}
M.L. Miller et~al.,
\newblock Ann. Rev. Nucl. Part. Sci. 57 (2007) 205.

\bibitem{Bossu:2011qe}
F. Bossu et~al.,
\newblock (2011), arXiv:1103.2394.

\bibitem{Book:2015}
J. Book,
\newblock Phenomenological interpolation of the inclusive J/psi cross section
  to proton-proton collisions at 2.76 TeV and 5.5 TeV (University of Frankfurt,
  2015).

\bibitem{Gavin:1988tw}
S. Gavin and M. Gyulassy,
\newblock Phys. Lett. B214 (1988) 241.

\bibitem{Hufner:1988wz}
J. Hufner, Y. Kurihara and H.J. Pirner,
\newblock Phys. Lett. B215 (1988) 218,
\newblock [Acta Phys. Slov.39,281(1989)].

\bibitem{Du:2018wsj}
X. Du and R. Rapp,
\newblock JHEP 03 (2019) 015.

\bibitem{Zhao:2007hh}
X. Zhao and R. Rapp,
\newblock Phys. Lett. B664 (2008) 253.

\bibitem{He:2013zua}
M. He et~al.,
\newblock Phys. Rev. E 88 (2013) 032138.

\bibitem{Cacciari:2012ny}
M. Cacciari et~al.,
\newblock JHEP 10 (2012) 137.

\bibitem{Laine:2006ns}
M. Laine et~al.,
\newblock JHEP 0703 (2007) 054.

\bibitem{Liu:2017qah}
S.Y.F. Liu and R. Rapp,
\newblock Phys. Rev. C 97 (2018) 034918.

\bibitem{Svetitsky:1987gq}
B. Svetitsky,
\newblock Phys. Rev. D 37 (1988) 2484.

\bibitem{Zhao:2010nk}
X. Zhao and R. Rapp,
\newblock Phys. Rev. C82 (2010) 064905.

\bibitem{Adam:2016rdg}
ALICE Collaboration, J. Adam et~al.,
\newblock Phys. Lett. B766 (2017) 212.

\bibitem{Akamatsu:2011se}
Y. Akamatsu and A. Rothkopf,
\newblock Phys. Rev. D85 (2012) 105011.

\bibitem{Borghini:2011ms}
N. Borghini and C. Gombeaud,
\newblock Eur. Phys. J. C72 (2012) 2000.

\bibitem{Brambilla:2016wgg}
N. Brambilla et~al.,
\newblock Phys. Rev. D 96 (2017) 034021.

\bibitem{Blaizot:2018oev}
J.P. Blaizot and M.A. Escobedo,
\newblock Phys. Rev. D98 (2018) 074007.

\bibitem{Katz:2015qja}
R. Katz and P.B. Gossiaux,
\newblock Annals Phys. 368 (2016) 267.

\bibitem{Chen:2016vha}
B. Chen, X. Du and R. Rapp,
\newblock Nucl. Part. Phys. Proc. 289-290 (2017) 475.

\bibitem{Yao:2018nmy}
X. Yao and T. Mehen,
\newblock Phys. Rev. D99 (2019) 096028.

\bibitem{Boyd:2019arx}
J. Boyd et~al.,
\newblock Phys. Rev. D 100 (2019) 076019, 1905.05676.

\bibitem{Brambilla:2020qwo}
N. Brambilla et~al.,
\newblock JHEP 05 (2021) 136, 2012.01240.

\end{thebibliography}

\end{document}